\documentclass[journal]{IEEEtran}
\usepackage{xcolor,soul,framed} 
\usepackage[pdftex]{graphicx}  
\usepackage[cmex10]{amsmath}   
\usepackage{array}             
\usepackage{mdwmath}           
\usepackage{mdwtab}            
\usepackage{eqparbox}          
\usepackage{url}               
\usepackage{xcolor}
\usepackage{cite}
\usepackage{hyperref}
\usepackage{graphicx}
\usepackage{booktabs}          

\colorlet{shadecolor}{yellow}

\begin{document}


\title{Computationally Efficient Spline-Based Modeling of DER Dynamics for Voltage Stability in Active Distribution Networks}
\author{
    Shadrack T. Asiedu, 
    Tara Aryal, 
    Zongjie Wang, 
    Hossein Moradi Rekabdarkolaee,
    Timothy M. Hansen
    \thanks{S. T. Asiedu and T. Aryal are with the Department of Electrical Engineering and Computer Science at South Dakota State University (SDSU), Brookings, South Dakota, USA 57007.}
    \thanks{Z. Wang and T. M. Hansen are with the Department of Electrical and Computer Engineering at Colorado State University, Fort Collins, Colorado, USA 80526. Corresponding author: tim.hansen@colostate.edu.}
    \thanks{H. Moradi Rekabdarkolaee is with the Applied Statistics and Operations Research Department at Bowling Green State University, Bowling Green, Ohio, USA 43403.}
    \thanks{This work is supported by the SDSU Richard McComish Ph.D. Fellowship in Energy Infrastructure, the U.S. Department of Energy Office of Science, Office of Electricity Microgrid R\&D Program, and Office of Energy Efficiency and Renewable Energy Solar Energy Technology Office under the EPSCoR grant number DE-SC0020281, and the U.S. National Science Foundation grant no. 2316400.}
}


\maketitle
\begin{abstract}
The increasing integration of Distributed Energy Resources (DERs) into power systems necessitates the accurate representation of their dynamic behavior at the transmission level. Traditional electromagnetic transient models (EMT), while effective, face scalability challenges due to their reliance on detailed system information. Data-driven approaches, such as System Identification (SysID), offer a promising alternative by modeling system dynamics without detailed system knowledge. However, SysID and similar methods are computationally intensive, requiring the computation of complex ordinary differential equations (ODEs) or transfer functions estimation. This makes them less effective for real-time operation. We therefore propose a novel data-driven approach that simplifies the modeling of DERs dynamics by leveraging B-splines to transform discrete system data into continuous differentiable functions. This enables the estimation of lower order linear ordinary differential equations with simple linear regression to represent the underlying dynamics at a very low computational cost. Furthermore, the extracted dynamic equations are discretized by the backward Euler method for potential integration into discrete-time power dispatch models.
Validation results indicate a goodness-of-fit (GoF) of 98.74\%, comparable to the 99.03\%  GoF of the SysID method, yet, 4.8 times faster. Our proposed model's execution time of less than one minute  makes it more suitable for real-time applications in power system operations.

\end{abstract}

\section{Introduction}
 Distribution systems have become more active with the increasing integration of distributed energy resources (DERs). These DERs such as solar PV, are connected to the grid via power electronic converters. Although they present the challenge  of low inertia, instability and harmonics, these inverter-based resources (IBRs) can also offer grid support functionalities to mitigate frequency and voltage deviations \cite{Poudel2024_Data_Driven_Modeling_of_Commercial_Inverters}. This requires accurate modeling of the DERs dynamics for real time coordination with the transmission system. Existing conventional models based on synchronous generators are too slow to capture the fast dynamics of  converter-dominated power systems. Popular methods used in literature for dynamic modeling of DERs include the switching models \cite{9446886}, averaged value models  \cite{Bountouris_2023_Dynamic_AVerage_Value_Model}, averaged linear models \cite{Wang2021_Small_Signal_Models_Overview}, dynamic phasor models \cite{Nagaragan_Dynamic_phasor}, and data-driven models \cite{Subedi_Data_Driven_Models_Review}. 
The switching models used for power system electromagnetic transient (EMT) simulations has been successfully implemented in \cite{Guo2024_EMT_Real_Time_Simulation} for dynamic studies of IBRs. The EMT simulation models every circuit element and semiconductor switch at their true time constants and solves the full set of differential algebraic equations in time domain. This makes the EMT method robust in capturing non-linearity and providing high accuracy. However, it suffers from high computational cost and the determination of many parameters \cite{Poudel2024_Data_Driven_Modeling_of_Commercial_Inverters}. As the system becomes larger with many high frequency converters, the computational cost becomes prohibitive, making their large scale application very difficult.

The average value model attempts to circumvent the high computational cost of the EMT model by averaging the detailed pulse width modulation switching waveform over each modulation period \cite{Jankovic2014_Average_Models}. To obtain an averaged linear model, the resulting model which is usually nonlinear, is linearized around an operating point to retain the key dynamics.\cite{Kwon_2017_Linearized_Modeling_Methods}. The feasibility of both the averaged value and averaged linear models for IBRs dynamics studies have been successfully demonstrated in \cite{Venkatramanan_2021_Average_Dynamic_Model_Three_Phase_Inverters,Bountouris_2023_Dynamic_AVerage_Value_Model}. Although the average models may reduce computational cost, they suffer the drawback of smoothing out some of the detailed system dynamics. This could impact the models' accuracy, making them less effective to study harmonic interactions between interconnected inverters \cite{Nazari2021_Dynamic_Phasor_Modeling_Three_Phase_Inverters}. Additionally, just like the EMT method, they require detailed knowledge of the system and the determination of a significant number of the system parameters. 

Dynamic phasor model avoids the detailed switching of EMT, and resolves the lack of explicit harmonic interactions of the average models by transforming each waveform into a set of slowly varying phasor for selected harmonics. This allows for larger time steps and explicit harmonic order analysis without full switching details. The method has successfully been tested in \cite{Koutenaei2022_Efficient_Dynamic_Phasor_Volt_Var} for dynamic modeling of power electronic converters. Although the dynamic phasor models offer better accuracy than the average models, and lower computational cost compared to the EMT methods, they still require knowledge about the harmonic content, control loops, circuit topology and parameters. 
An advanced EMT and phasor-domain hybrid simulation approach has also been proposed in ~\cite{huang2018advanced}, offering higher accuracy. However, aside from requiring knowledge of the system, this approach introduces mathematical complexity, making small signal stability analysis more challenging ~\cite{shah2021review}. These limitations make data-driven approaches preferable for black-box modeling of DERs dynamics as little to no system knowledge is required.

One of the most popular data-driven methods for dynamic modeling is the system identification SysID technique, utilized in \cite{guruwacharya2022data,Subedi2024Aggregate,Sunil_20221_partition_method} for IBRs dynamic modeling. By extracting time invariant operational or simulated dataset from the system, SysID estimates the gains and time constants of transfer functions to represent the system dynamics. Although the SysID method has proven to be effective for dynamic modeling, its computational cost still remains significantly high, especially with increasing order of the transfer functions. This is largely due to the non-linear programming problem formulation and the use of complex optimization methods to find the optimal coefficients.

To address this problem, our study significantly reduces this computational burden by fitting a differentiable smooth function to the dataset and applying linear regression to extract the system dynamics in the form of ordinary differential equations (ODEs). By smoothing the data to remove noise, it ensures the underlying dynamics of the system can easily be captured with simple models such as linear regression.  The main contributions of this study are:
\begin{itemize}
    \item Demonstrated the use of B-splines to fit smooth and differentiable curve to dynamic system dataset.
    \item Extracted and discretized a set of ODEs from the smooth curve to represent the aggregate dynamics of the DERs.
    \item Compared the computational cost and scalability of the proposed model with the SysID technique.
\end{itemize}

The rest of the paper is organized as follows. Section~\ref{sec:Design} presents a description of how the system under study was designed in simulink. The theoretical and mathematical formulations of the proposed model are presented in Section~\ref{sec:Dynamic_Modeling}. The simulation setup is shown in Section~\ref{sec:Simulation_Step}, with a presentation of results and discussions in Section~\ref{sec:Results}. Finally, conclusions are drawn in Section~\ref{sec:Conclusion}.

\section{Design of the Dynamic System} \label{sec:Design}

\subsection{Grid Support Power Electronics Converters}
\begin{figure}[ht!]
    \centering
    \includegraphics[width=3.4in]{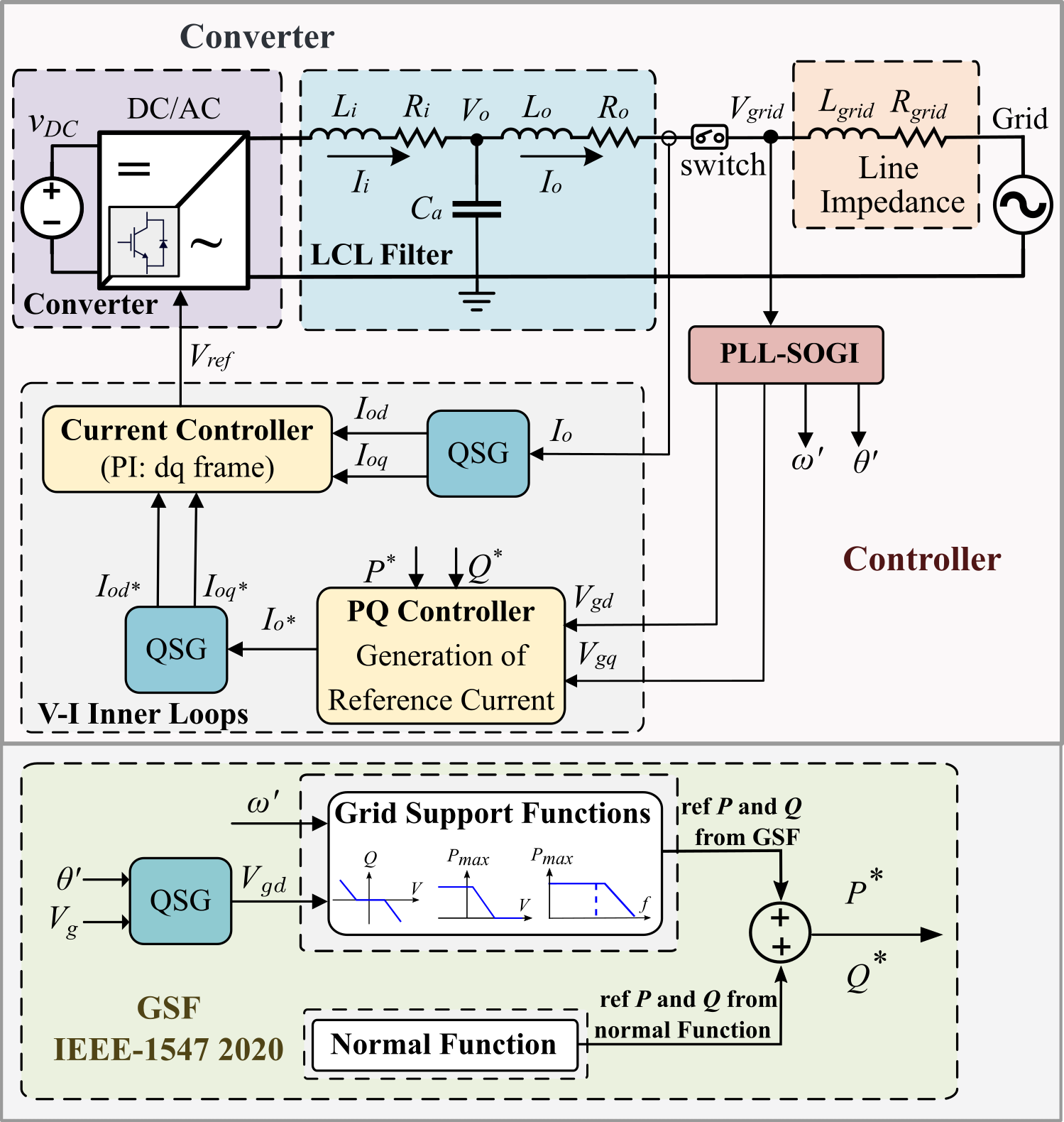}
    \caption{Illustration of typical grid-connected 1 $\phi$ power electronics converter with grid support functions system \cite{Subedi2024Aggregate}}
    \label{fig:GSF}
\end{figure}

In the Fig.~\ref{fig:GSF}, a single-phase, single-stage voltage source power electronics converter (PEC) with grid support functions (GSFs) is connected to the primary side of the grid through an LCL filter \cite{Subedi2024Aggregate}. The LCL filter consists of an inverter-side inductor ($L_i$), a capacitor ($C_0$), and a grid-side inductor ($L_0$), ensuring a high-quality sinusoidal grid current. In this configuration, a single-stage system without a DC-DC converter is employed, assuming the DC voltage has already been boosted. The key variables $V_c, I_i, V_g$ and $I_0$ denote the voltage of a capacitor, input current, voltage of the grid, and the output current, respectively. The output current ($I_0$) to the primary grid is regulated by a proportional-integrator type 2 (PI-2) controller. The magnitudes of $I_0$ and $V_g$ injected by the PEC are measured using a second-order generalized integrator (SOGI), based on a quadrature signal generator (QSG). This method also measures the phase angle ($\theta'$) and the fundamental frequency ($\omega'$).

The PEC can switch between grid-supporting and grid-feeding modes based on the operational conditions. The reference current ($I_0^*$) is generated by the PQ power controller, which relies on the active power ($P^*$) and reactive power ($Q^*$). The reference power values, $P^*$ and $Q^*$, depend on the type of inverter used. For instance, smart inverters generate reference power based on grid frequency and voltage, while traditional inverters provide a constant maximum power output. This reference current is then fed into the PI-2 controller, as shown in the Fig.~\ref{fig:GSF}

\subsection{Probing Signal}
To accurately extract system information and understand its behavior, it is essential to perturb the system using a well-designed probing signal. The probing signal must be capable of capturing a wide range of system dynamics, including various operating modes. According to the classical approach outlined in ~\cite{pierre2009probing}, several key considerations should guide the design of such signals. One important criterion is the power content of the input signal. This needs to be optimized to focus the frequency spectrum within the desired dynamic range of the model to enhance the model accuracy. Furthermore, the amplitude of the input signal should span the full range between the system’s minimum and maximum voltage limits so as to ensure optimal signal strength and improve the signal-to-noise ratio.

A review of the literature indicates that the logarithmic square chirp signal outperforms other probing signals—such as pure sine, logarithmic sine, and square waves—in terms of effectively capturing power system dynamics ~\cite{guruwacharya2022data, rauniyar2021evaluation}. The logarithmic square chirp signal combines a square wave with a logarithmic frequency sweep, thereby offering a rich frequency content suitable for system identification. Its design takes into account the principles of system identification theory and the practical constraints of power system operation, including amplitude limits, signal energy, and time duration.

The fundamental equation of the probing signal used in this research is given below in (\ref{eqn:chirp}):

\begin{equation}
    x(t)=Acos(2\pi f(t) + \phi_0)
    \label{eqn:chirp}
\end{equation}

Here, $A$ represents the amplitude of the signal, $f(t)$ denotes the instantaneous frequency at time $t$ and $\phi _0$ is the initial phase angle. The frequency profile of the chirp signal is defined by three parameters: the starting frequency $f_0$, the ending frequency $f_1$, and the total duration of the chirp signal $T$. The time-varying frequency sweep $f(t)$ is mathematically defined as:
\begin{equation}
    f(t)=f_0 \times \frac{f_0(\frac{1}{T})^t}{f_1}
    \label{eqn:freq}
\end{equation}
 The ending frequency $f_1$ is selected based on the system's time constant, which can be estimated by analyzing the settling time of the system's step response. The starting frequency $f_0$ is typically chosen as a fraction of $f_1$, commonly set to $\frac{f_1}{100}$. Once $f_0$ and $f_1$ are determined, the duration of the chirp signal $T$ can be calculated using (\ref{eqn:freq}).

\section{Dynamic Modeling of DERs} \label{sec:Dynamic_Modeling}
\begin{figure}[ht!]
    \centering
    \includegraphics[width=3.4in]{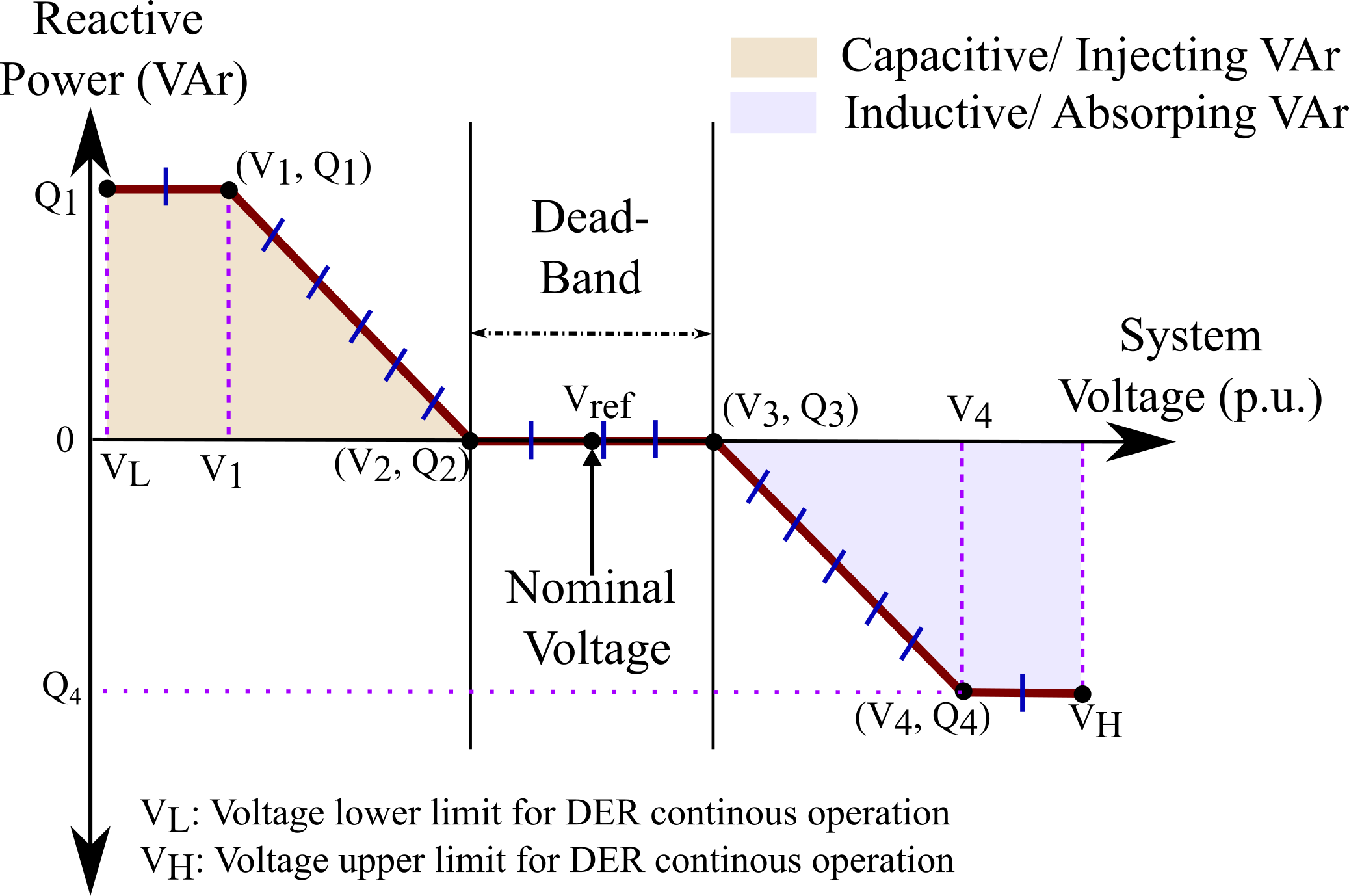}
    \caption{Partitioned IEEE 1547-2020 Volt-Var Curve \cite{Sunil_20221_partition_method}}
    \label{fig:Volt-Var}
\end{figure}

The dynamic modeling of DERs in this study is the continuous time representation of the inverter's reactive current response to changes in voltage at the point of common coupling (POC). The DERs, equipped with Volt-Var grid support function is meant to maintain voltage stability at the POC. The Volt-Var operation of the DERs inverters are set in accordance to the IEEE 1547-2018 standard curve, which is a piecewise linear function \cite{IEEE_1547}. This piecewise linear curve defines different voltage bands where the inverter injects or absorbs reactive current, when the voltage moves outside the normal operating range. However, in reality, the piecewise curve presented by the IEEE 1547-2018 is not perfectly linear, presenting modeling challenges. To resolve this, the partitioned model has been used in \cite{Sunil_20221_partition_method, Guruwachargy_HIL-Inverer_2022} to further divide the piecewise linear curve into further smaller linear intervals. Each interval is then represented by a more simplified linearized equation. Fig. \ref{fig:Volt-Var} shows the partitioned Volt-Var curve adopted for this study.

\subsection{System Identification Model}
The System Identification (SyID) technique is used to build mathematical models from measured data. Reduced order dynamic linear transfer functions (RODLTF) \cite{SUBEDI2023100365} have become a popular model structure to represent the dynamic systems. Transfer functions relates the output of a system to its input in the Laplace domain (for continuous systems) or the Z-domain (for discrete systems) \cite{Kang2006_s-to-z_transfer_function}, to characterize the dynamic behavior of linear time-invariant (LTI) systems. Based on the input and output data from the system, the coefficients of the transfer functions are estimated  by minimizing the error between the estimated output and the measured output of the system. Generally, a transfer function is a ratio of polynomials for both the continuous and discrete system. For a discrete system , which is used in this study, the $Z$-transform that gives the transfer function is defined as:

\begin{align}
Y(z) &= \frac{b_0 + b_1 z^{-1} + b_2 z^{-2} + \cdots + b_m z^{-m}}{1 + a_1 z^{-1} + a_2 z^{-2} + \cdots + a_n z^{-n}} U(z) \\ \nonumber
&+ M(y_0, y_1, \dots, y_{n-1})
\end{align}

where \( m \) and \( n \) are the order of the numerator and denominator, respectively,  \( [a_1, a_2, \dots, a_n] \) and \( [b_0, b_1, \dots, b_m] \) are the coefficients, and \( M(y_0, y_1, \dots, y_{n-1}) \) is an initial condition term.

\subsection{Functional Data Analysis (FDA)}
Functional data analysis (FDA) provides a means to transform discrete data into continuous functions. Using basis functions such as Fourier basis for periodic data and B-splines for non-periodic data, a noisy high dimensional data could be streamlined into a single smooth function for easy analysis \cite{Ramsay2005_FDA}. The two popular FDA modeling approaches are functional linear regression, which fits a smooth curve to discrete data, and functional principal component analysis, which reduces a set of functional data into a few principal component functions \cite{Wang2015_FDA_Review}. For this study, we adopt the functional linear regression using B-splines as the basis functions.

\subsubsection{B-Splines}
B-splines are non-negative piecewise polynomial joined end-to-end smoothly at points known as knot. The number of knots, which forms the grid size, together with the order of the splines, determines the number of B-spline basis functions. Thus, the number of basis functions ($nbasis$) is determined by :
\begin{align}
nbasis = order + number\ of\ knots - 2 
\end{align}
In instances where the grid size and degree of the polynomial are rather specified, the number of basis functions is obtained as:
 
\begin{align}
nbasis &= degree + grid\ size
\end{align}

\begin{figure}[ht!]
    \centering
    \includegraphics[width=3.4in]{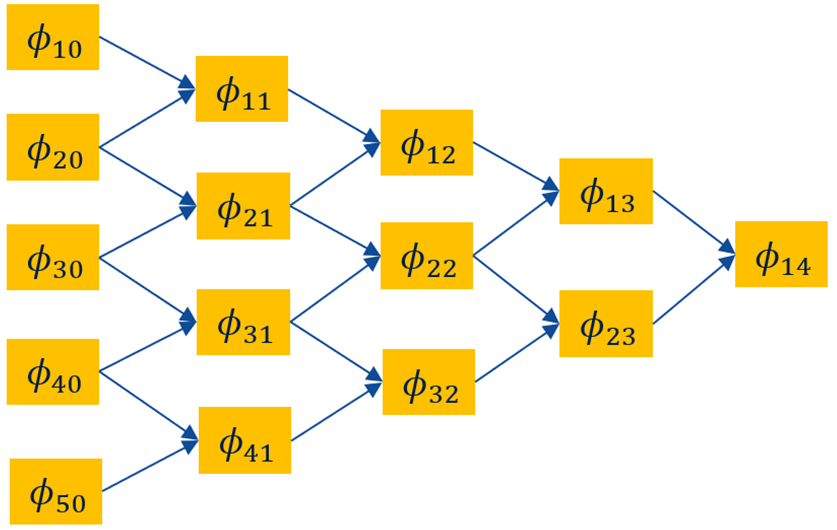}
    \caption{Recursive formulation of A Fourth Order Bspline}
    \label{fig:Bspline_basis}
\end{figure}

As could be seen from Fig.~\ref{fig:Bspline_basis}, each basis function $(\phi_{i,n})$ of order $n$ is obtained recursively from two preceding basis functions of order $n-1$. 

\begin{figure}[ht!]
    \centering
    \includegraphics[width=3.4in]{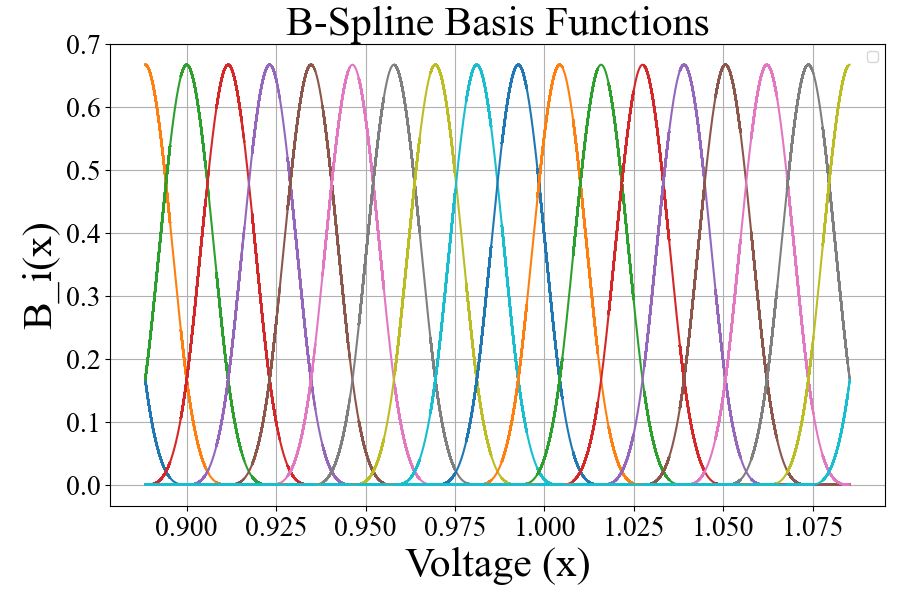}
    \caption{20 Cubic B-spline Basis Functions}
    \label{fig:Bspline_basis_plot}
\end{figure}

The recursive formulation is derived from the Cox-de Boor equation which is given as \cite{Cox1972_Bspline}:
\vspace{-5pt}
\begin{align}
\phi_{0,j}(x) &= 
\begin{cases} 
1 & \text{if } t_1 \leq x < t_{i+1} \\
0 & \text{otherwise}
\end{cases} \\
\vspace{-5pt} 
\phi_{i,j+1}(x) &= \alpha_{i,j+1}(x) \phi_{i,j}(x) + [1 - \alpha_{i+1,j+1}(x)] \phi_{i+1,j}(x)] \\
\vspace{-5pt} 
\alpha_{(i,j)}(x) &= 
\begin{cases} 
\frac{x - t_1}{t_{i+j} - t_i} & \text{if } t_i + j \neq t_1 \\
0 & \text{otherwise}
\end{cases}
\end{align}

where \( x \) is the time point, \( t \) is the knot sequence, and \( \phi_{i,j} \) is the \( i \)-th basis function of degree \( j \).
Different order of B-splines produces different representation of the discrete data. A second order B-splines are most suitable for dataset with linear characteristics. However, where there are curvatures, higher orders are most adequate. As a general rule of thumb, the order of the spline should be at least $k + 1$ if interested in $k$ derivatives. Due to this the cubic B-spline is most popularly used since many dynamic system analysis are focused on the gradient and the rate of change of the gradient. Additionally, until explicitly specified, the knots of the splines are evenly spaced by default. Changing the number of knots or the order of  splines automatically changes the number of basis functions. A smaller number of basis functions results in little flexibility, while a larger number of basis functions increases flexibility, allowing the model to capture detailed patterns. However, too many basis functions may overfit, making the number of basis function a key hyperparameter. Further, the range of the knots should always span the range of the dataset in order to adequately transform the discrete data into function. For instance, Fig.~\ref{fig:Bspline_basis_plot} shows the 20 basis functions used for our study, spanning the entire voltage range from  0.88-1.10pu.

\subsubsection{Data to Function Transformation}
 Given a dataset of single input $x_i$, and single output $y_i$, if we represent the basis function as $\phi_j$,  the general functional representation of the dataset can be written as \cite{cao2019github} :
\begin{align}
y_i = \sum_{j} c_j \phi_j (x_i) + \epsilon_i = f(x_i) + \epsilon_i,
\end{align}

\text{thus, in vectorized form:}

\begin{align}
f(x) = c^T \phi(x)
\end{align}

\text{where:}

$\phi(x)$ is the basis function, $\epsilon$ is the estimation error, and $c$ is a vector of the coefficients.
Using least squared estimation method, the analytical expression of the model could be obtained by: 
\begin{align}
H &= \sum_{i=1}^n (y_i - f(x_i))^2 &  \\
  &= \sum_{i=1}^n (y_i - c^T \phi(x))^2 & 
\end{align}

\text{using vectorization:}
\begin{align}
H &= (\mathbf{y} - \mathbf{\Phi c})^T (\mathbf{y} - \mathbf{\Phi} \mathbf{c}), & 
\end{align}

\text{the sum of square errors is minimized to obtain}
\begin{align}
\hat{c} &= (\mathbf{\Phi}^T \mathbf{\Phi})^{-1} \mathbf{\Phi}^T \mathbf{y} \,, \label{eq:least_squares}
\end{align}

\text{the estimated function is then given as:}
\begin{align}
\hat{y} &= f(x) = c^T \mathbf{\Phi}(x) & \label{eq:estimated_model} 
\end{align}

\text{where \( H \) is the loss function and \( \hat{y} \) is the estimated model.}

By substituting (\ref{eq:least_squares}) into (\ref{eq:estimated_model}), we obtain:
\begin{align}
\hat{\mathbf{y}} &= f(x)= \boldsymbol{\Phi}(x)(\boldsymbol{\Phi}^T \boldsymbol{\Phi})^{-1} \boldsymbol{\Phi}^T \mathbf{y} &  \label{eq:fda_equation}
\end{align}

\begin{align}
\hat{\mathbf{y}} &= \mathbf{S} \mathbf{y} &
\end{align}

where $\mathbf{S}$ is the smoothing matrix. The trace of $\mathbf{S}$ gives the number of basis functions and number of coefficients.

\subsubsection{Smoothing Splines}
The core application of functional data analysis is to generate a smooth function  with a continuous derivatives. This, however, is not easily attainable by the initial representation of the discrete data with parameterized basis functions. The typical B-splines representation focus on goodness of fit to  represent the discrete data with a continuous function. In many cases, this results in continuous functions which are not perfectly smooth, especially with the wrong selection of knots and order of the splines. The conventional way to resolve this to tune the number of knots. However, fewer number of knots may increase the smoothing but reduce the goodness of fit. Conversely, a large number of knots could also increase the goodness of fit but significantly affect the smoothness of the curve. This situation precipitates the introduction of a smoothing parameter $(\lambda)$ into the optimization problem to strikes a good balance between smoothness and goodness of fit. Thus, instead of simply using a least squared estimation method, a penalized least squared estimation method is used to smoothen the results obtained in (\ref{eq:fda_equation}). That is, a $\lambda$ is obtained such that:

\begin{align}
\min_{}   J_{\text{smooth}}(\lambda) = \left[\mathbf{y} - \hat{\mathbf{y}}(x)\right]^T \left[\mathbf{y} - \hat{\mathbf{y}}(x)\right] + \lambda J_m (x)
\end{align}

In this context, $J_m$ denotes the integral of the square of the \( m \)-th derivative of the estimated model defined in (\ref{eq:fda_equation}).
Penalizing this term ensures a satisfactory level of smoothness is achieved without overly compromising the goodness of fit. $J_m$ is then given as:
\begin{align}
J_m (x) = \int [D^m \boldsymbol{\hat{y}(x)]^2} \, dx
\end{align}

The value of $m$, which is the order of the desired differential analysis, heavily influences the the order $(k)$ of the B-splines. Where second derivative is desirable, a fourth order B-spline is necessary for the transformation of the discrete data into function. 
Assuming a second derivative analysis in our study, we select $m=2$ such that:

\begin{align}
\int \left[ D^2 \mathbf{\hat{y}}(x) \right]^2 \, dx 
= \int \mathbf{c}^T \, D^2 \mathbf{\Phi}(x) \, \left( D^2 \mathbf{\Phi}(x) \right)^T \mathbf{c} \, dx
\end{align}

For simplification, let

\begin{align}
R = D^2 \mathbf{\Phi}(x) D^2 \mathbf{\Phi}^T,
\end{align}
such that:
\begin{align}
R_{\text{mat}} = \mathbf{c}^T R
\end{align}
where \(R\) is called the penalty matrix.

The penalized least square estimate of the coefficients finally becomes
\begin{align}
\hat{\mathbf{c}}_p = (\mathbf{\Phi}^T \mathbf{\Phi} + \lambda R_{\text{mat}})^{-1} \mathbf{\Phi}^T \mathbf{y}
\end{align}

The final smoothed curve is estimated as:
\begin{align}
\mathbf{\hat{y}_p} =f_p(x) = \mathbf{\hat{c}}_p^T \boldsymbol{\Phi(x)}
\end{align}

Conventionally, the local optimal value of the smoothing coefficient ($\lambda$) could be obtained by iterating through a neighborhood ($N$) of manually selected possible values of lambda ($\lambda$). However, one of the most effective means of obtaining the optimal value of $\lambda$ is by using the Ordinary Cross Validation (OCV) method such that:

\begin{align}
\text{OCV}(\lambda) &= \frac{1}{n} \sum \left( \frac{y_i - \hat{y}(x_i)}{1 - S_{ii}} \right)^2 &
\end{align}

where \( S \) is the smoothing matrix.

\subsection{Ordinary Differential Equations (ODEs)}

Ordinary Differential Equations (ODEs) are crucial for describing system dynamics in various fields of engineering and sciences. These equations define the relationships between different states of a system and their rates of change over time. ODEs are particularly significant in power systems, where they model the dynamic behaviors of components such as generators and motors. The representation through ODEs allows for the analysis of system responses to external inputs and disturbances, crucial for stability analysis and control measures. 
An ODE typically involves derivatives of a function and can be classified by its order. The order of an ODE is determined by the highest derivative it contains. The general form of an ODE is expressed as:

\begin{align}
a_n(t)\frac{d^n y}{dt^n} + a_{n-1}(t)\frac{d^{n-1} y}{dt^{n-1}} + \ldots + a_1(t)\frac{dy}{dt} + a_0(t)y = g(t), &
\end{align}
where \(a_i(t)\) are the coefficients and \(g(t)\) represents the driving function.
One of the primary challenges in solving ODEs is the necessity for initial conditions or guesses, which is critical in non-linear dynamics where the system's behavior can diverge significantly based on initial values. This sensitivity, in addition to the complex optimization of higher order ODEs, makes modeling and simulation particularly challenging.

However, this problem could be resolve if the derivatives of the underlying functions of the dataset are easily determinable. This idea forms the basis of our novel approach, resolving the associated complexities. This is achieved by fist transforming discrete data into smooth functions. The smoothness of the function makes it continuously differentiable to obtain the rate of change. With the input and output variables known, a simple linear regression model could be used to estimate coefficients of a linear ODE that accurately represents the system dynamics.
A simple first order linear ODE is first experimented to prove this concept, such that:

\begin{align}
 I(t) &= AV(t) + B\frac{dI}{dt} + C &, \label{eq:regress_analysis} 
\end{align}

If a  second order ODE is required, we can have, 

\begin{align}
 I(t) &= AV(t) + B\frac{dI}{dt} + C\frac{dI^2}{d^2t} + D , &
\end{align}

where \( I \) is the measured aggregate current response of the inverters, \( V \) is the observed voltage at the point of interconnection, and \( A \), \( B \), \( C \) and \(D\) are the coefficients to be estimated. The system dynamics is modeled such that the current response is dependent on the voltage magnitude, the rate of change of the current and an offset value. This linear model is estimated for each of the 20 partitions of the model.

\subsection{Discretization of ODE}
The primary objective of modeling power system dynamics is to incorporate the derived behaviors into comprehensive power system models. While ordinary differential equations (ODEs) describe system dynamics through continuous functions, discretization of these ODEs is crucial. It converts the continuous system dynamics into a discrete-time format, which aligns with the timestamped operational requirements of applications like power dispatch. This transformation is essential for integrating dynamic models seamlessly into time-stepped processes of power system operations. Four major discretization methods have been discussed in \cite{arredondo2019analysis}. The Forward Euler and the Runge Kutta methods are explicit methods that basically predict future value by approximating the derivative at the current timestep. The Runge Kutta method, provides higher accuracy by averaging the derivatives at multiple intermediate steps within the current time step. In contrast, the implicit methods such as the Backward Euler and Trapezoidal rule, consider the derivative at the next time step to update the current value \cite{Rapp2017}. The Trapezoidal rule however, averages the derivatives at both the current and future time steps to enhance the prediction accuracy. While each discretization method has its strength and weakness, the Backward Euler method is chosen for this study due to its greater stability against stiff problems. The generic formulation of the Backward Euler method is given as \cite{arredondo2019analysis}:

\begin{align}
y_{k+1} = y_k + \Delta t \cdot F_d(y_{k+1}, x_{k+1}, t_{k+1}), \label{eq: backward_euler}
\end{align}
where $y_{k+1}$ is the state variable at the next time step, $F_d(y_{k+1}, x_{k+1}, t_{k+1})$ represents the differential equation evaluated at the next time step.
To discrete the extracted linear dynamic equations, (\ref{eq:regress_analysis}) is rewritten such that:

\begin{align}
\frac{dI}{dt} = \frac{1}{B} \left(I(t) - AV(t) - C\right) = F_k(I, V, t)
\end{align}

Assuming $I_k$ and $I_{k+1}$ as the current state and the next state respectively, then by applying the Backward Euler method to the extracted linear ODE gives:

\begin{align}
I_{k+1} &= I_k + \frac{\Delta t}{B} (I_{k+1} - AV_{k+1} - C) \\
I_{k+1} - \frac{\Delta t}{B} I_{k+1} &= I_k - \frac{\Delta t}{B} (AV_{k+1} + C) \\
\left(1 - \frac{\Delta t}{B}\right) I_{k+1} &= I_k - \frac{\Delta t}{B} (AV_{k+1} + C)
\end{align}
By letting \( D = \frac{\Delta t}{B} \), the final discretized ODE representing the system dynamics is obtained as:
\begin{align}
I_{k+1} = \frac{1}{1 - D}  I_k - \frac{D}{1 - D} (AV_{k+1} + C), \label{eq:backward_euler_eqn} 
\end{align}

\section{Simulation Setup} \label{sec:Simulation_Step}
\begin{figure}[ht!]
    \centering
    \includegraphics[width=3.7in]{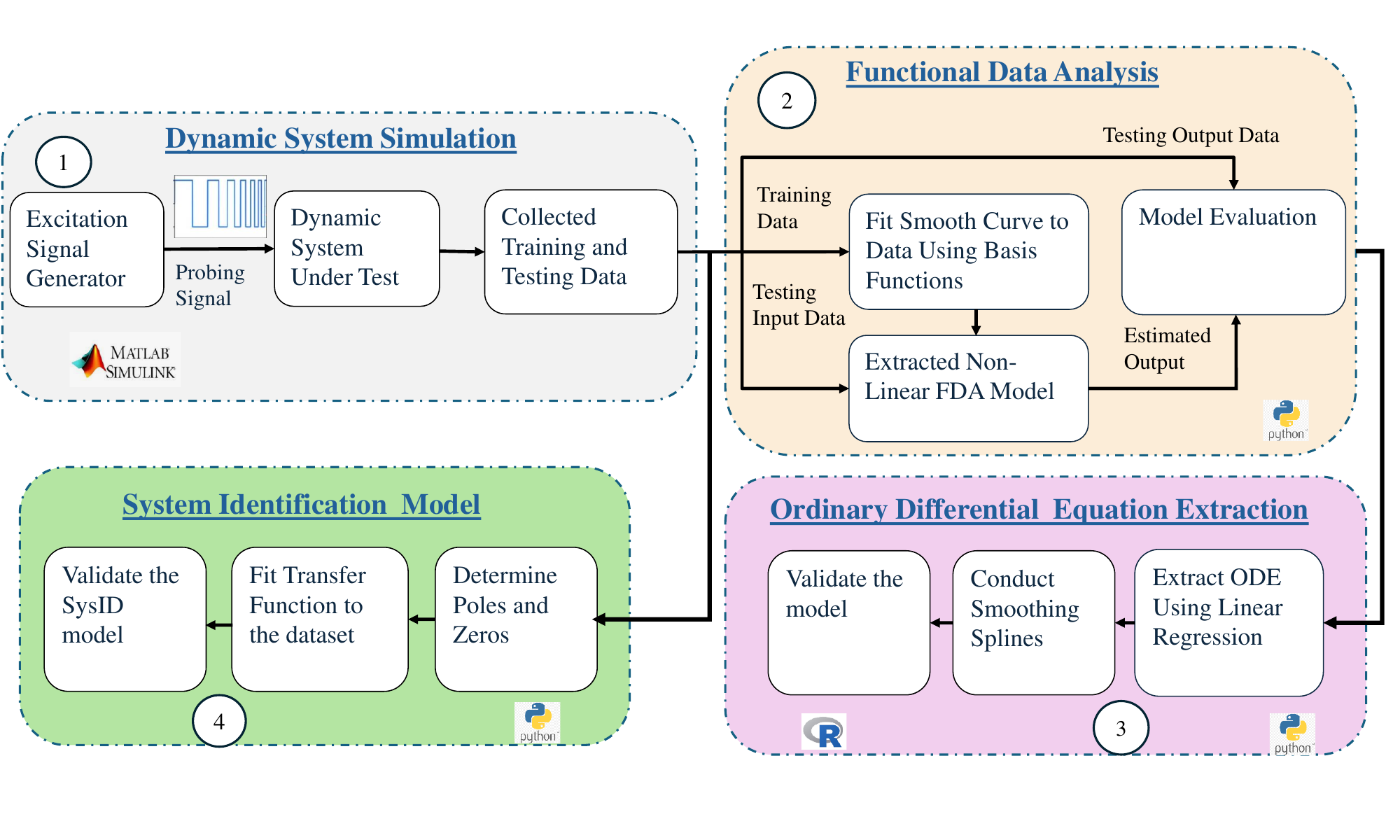}
    \caption{Proposed Framework of the Bspline-Based Dynamic Model}
    \label{fig:Model_Framework}
\end{figure}

In study, we excited a 36 house distribution system with probing signal to measure the aggregate reactive current response of the inverters connected to solar PV at each house. Using functional data analysis, the extracted dataset is transformed into smooth and continuous function. The time derivative of the current is determined and the dataset is divided into 20 partitions. We then proceeded to fit a linear ODE for each partition using least squared estimation to determine the coefficient of the linear models.
The model is validated using square step signal, and the performance compared to SysID model on the same datasets. The models were developed in Python version 3.11.7 and R version 4.3.2 running on a local desktop equipped with an Intel\textsuperscript{\textregistered} Core\textsuperscript{\texttrademark} processor featuring 16 CPUs, each operating at a clock speed of 2.0 GHz and 16 GB of RAM. The methodological framework is shown in Fig.~\ref{fig:Model_Framework}.

\subsection{Data Collection and Analysis}
\begin{figure}[ht!]
    \centering
    \includegraphics[width=3.4in]{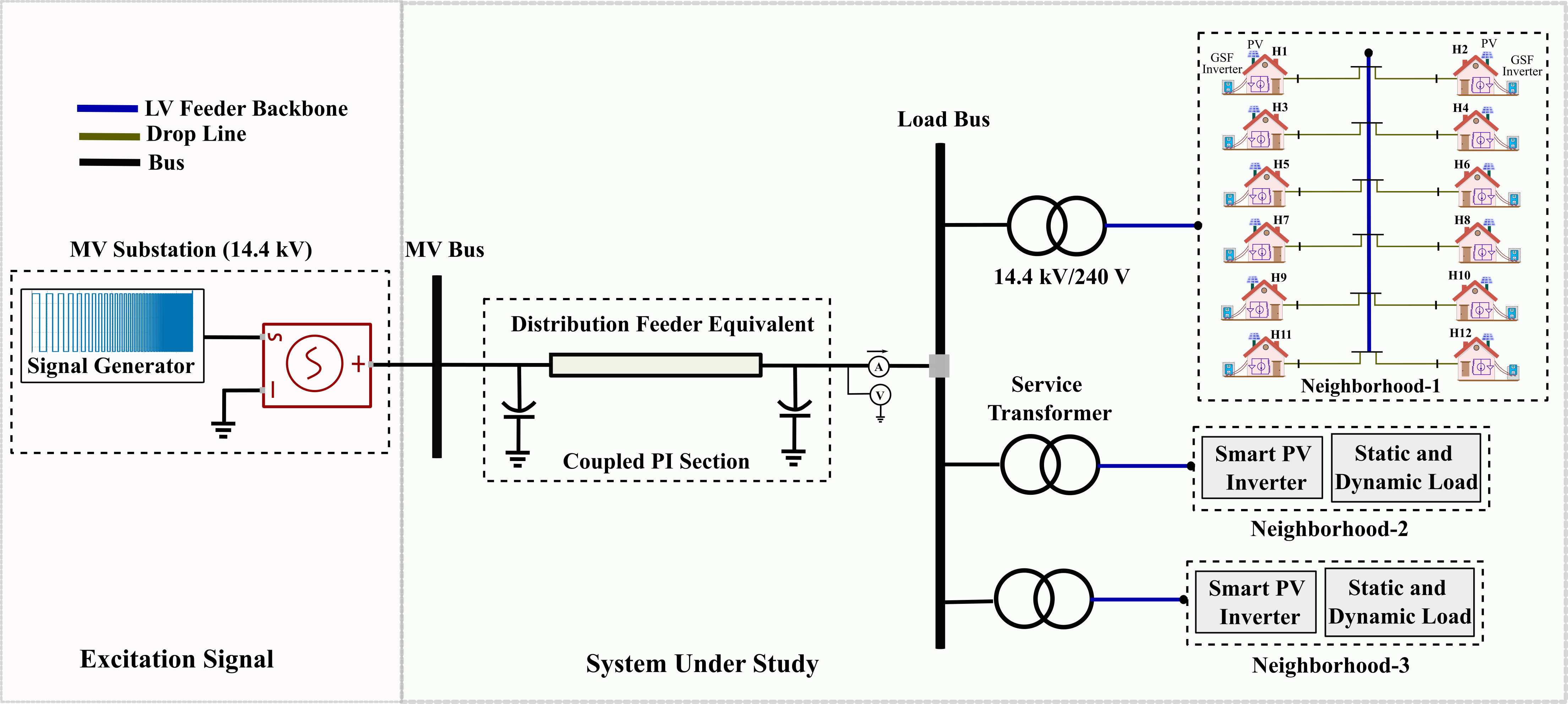}
    \caption{Medium Voltage substation  and feeder line connected to 1 $\phi$ low voltage distribution network neighborhood system with CMLD and DER inverters \cite{Sunil_20221_partition_method}}
    \label{fig:DN}
\end{figure}

\begin{figure}[ht!]
    \centering
    \includegraphics[width=3.4in]{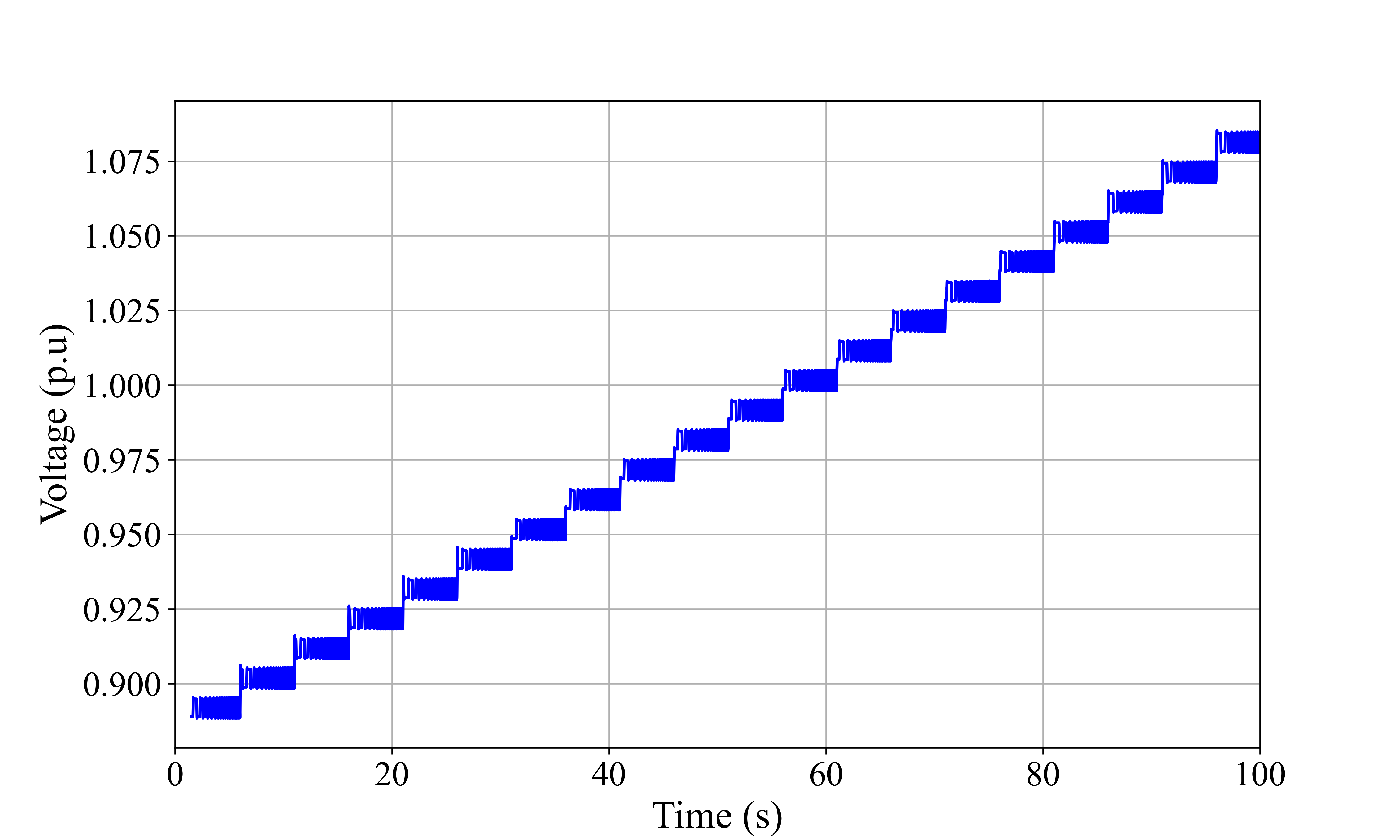}
    \caption{Logarithm Square Chirp Signal for Excitation (Voltage Variation)}
    \label{fig:Volt_Profile}
\end{figure}

\begin{figure}[ht!]
    \centering
    \includegraphics[width=3.4in]{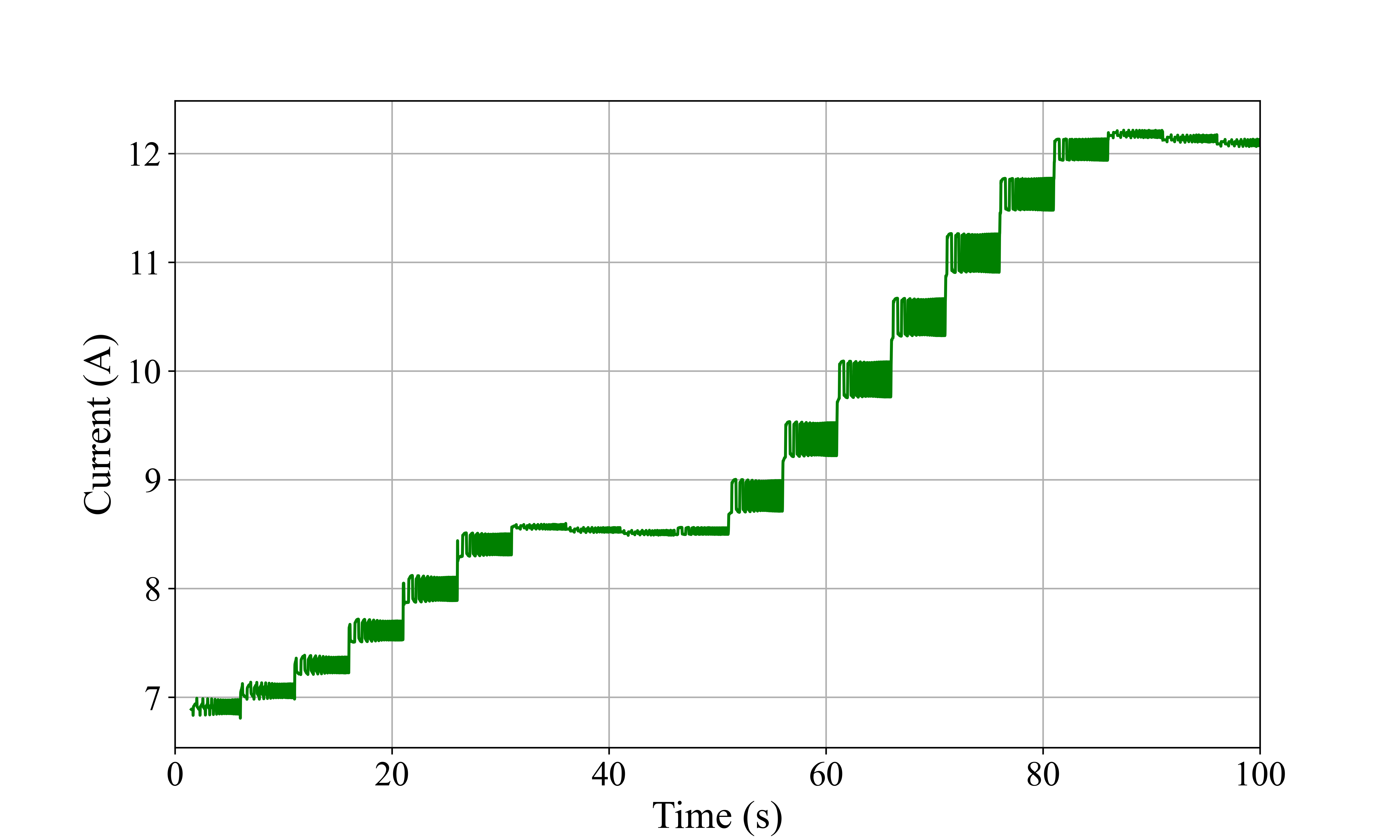}
    \caption{Profile of Aggregate Current Response of DERs}
    \label{fig:Curr_Profile}
\end{figure}

The DERs dominated distribution system used for this study is shown in Fig.~\ref{fig:DN}. It is composed of a 36 houses, each house equipped with a solar PV system connected to the grid through an inverter operating in a Volt-Var mode. Each house serves a composite load (CMLD).
The data  was obtained by  exciting the system in Fig.~\ref{fig:DN} with a logarithmic square chirp signal for 100 seconds. The values of the input signal ($V$), the timesteps ($t$), and the aggregate current response ($I$) of the DERs' inverters were extracted as the training data. Fig.~\ref{fig:Volt_Profile} and Fig.~\ref{fig:Curr_Profile} show the profile of the voltage and current used for training the models. Details of the parameters, state variables of the probing signal, and the inverter settings are presented in Table~\ref{tab:System_Parameters}.

\begin{table}[h]
\centering
\caption{Inverter and probing signal design parameters.}
\label{tab:inverter_params}
\begin{tabular}{@{}lc@{}}
\toprule
\textbf{Parameter}                         & \textbf{Value}                \\ \midrule
\multicolumn{2}{@{}l@{}}{\textbf{Inverter parameter}}                  \\
Rated capacity                             & 6.25 kW                       \\
Net-power                                  & 4.375 kW                      \\
$L_i, R_i$                                 & 0.1$\Omega$, 0.0018H          \\
$L_o, R_o$                                 & 0.1$\Omega$, 0.0018H          \\
$C_a$                                      & 8.8 $\mu$F                    \\ \midrule
\multicolumn{2}{@{}l@{}}{\textbf{Current PI-2 controller parameter}}   \\
$K_P$                                      & 4.5                           \\
$K_I$                                      & 60                            \\ \midrule
\multicolumn{2}{@{}l@{}}{\textbf{Volt-VAr setting}}                  \\
$Q_1, Q_2, Q_3, Q_4$                       & 4.375, 0, 0, -4.375 kW        \\
$V_L, V_1, V_2$                            & 0.88, 0.92, 0.98 p.u.         \\
$V_3, V_4, V_H$                            & 1.02, 1.08, 1.1 p.u.          \\ \midrule
\multicolumn{2}{@{}l@{}}{\textbf{Probing signal parameter}}            \\
Amplitude (A)                              & 0.8884–1.0884 p.u.            \\
Frequency ($f_0$ to $f_1$)                 & 1–5 Hz                        \\
Chirp sweep time (T)                       & 5 s                           \\
Rate of increase in frequency ($\lambda$)  & 1\%                           \\ \bottomrule
\end{tabular}
\label{tab:System_Parameters}
\end{table}

Additionally , the same system is excited for 50 seconds with square step signal with amplitude variation from 0.90pu to 0.99pu. The extracted values of the voltage, timesteps and current were used to test the performance of the two models. For each round of excitation, the initial 1.20 seconds show transient disturbances resulting from the initial connection of the inverters to the grid. Therefore, both the training and testing datasets were cleaned of the first 1.20 seconds data points.

\subsection{FDA Modeling}
The FDA approach in this study leverages B-spline basis functions to transform discrete system data into continuously smooth functions. This is to  enable the computation of derivatives necessary for dynamic modeling. The first step involves converting the voltage and current data into PyTorch tensors. This is to to ensure they align with the computational framework of the \textit{$B\_batch$} function from \textit{kan.spline} library in python developed by \cite{liu2025kankolmogorovarnoldnetworks}. The B-splines were created with a grid size of seventeen (17) and spline order of three (3). This gave a total of twenty (20) B-spline basis functions for the FDA modeling. The {$B\_batch$} function adopts the ordinary cross validation method to estimate the optimal value of the smoothing coefficient ($\lambda$) which gives a good balance between accuracy and smoothness.

 Each point in the dataset is mapped onto this spline basis, effectively transforming the discrete data points into a smooth and continuous function representation. The B-splines' coefficients are then calculated using least squared estimation given in (\ref{eq:least_squares}). The resulting model, developed using (\ref{eq:fda_equation}), provides a smooth approximation of the dataset, offering insights into the continuous behavior of the system without the computational complexities of traditional numerical methods.

\subsection{Extraction of First Order ODEs}
Following the development of the smooth function representation of the dataset, the next phase involves the regression-based extraction of a set of ODEs that governs the system dynamics using (\ref{eq:regress_analysis}). The differential aspect of the system’s dynamics is captured by computing the first derivative of the current with respect to time, to serve as the  pivotal component of the dynamic model. Traditionally, performing this derivative on the discrete data without the smooth function transformation would have produced some infinite values, making it impossible to proceed with the dynamic modeling.

The dataset is further partitioned into 20 segments based on voltage intervals. Each segment represented a voltage range of 0.01pu. The optimal number of partitioned were obtained by experimenting with different number of partitions from 5 to 30. Each of the selected 20 partitions undergoes a separate linear regression analysis to model the relationship between the voltage, current, and their temporal derivatives. This partition approach does not only enhances the accuracy of the model by adapting to local characteristics of the data, but also, simplifies the model by focusing on linear relationships within each voltage range. The regression coefficients obtained from each segment provide a detailed characterization of the system’s dynamics. Finally, we aggregated the predicted current responses of all the segments to give the overall dynamic response of the system.

\subsection{Discretization of Dynamic Equation}
The final step in our proposed model involves discretizing the extracted ODE model using the Backward Euler method in (\ref{eq:backward_euler_eqn}). This approach is very important for the integration of the dynamic model into discrete-time  power system operation such optimal power flow. The discretization involves the reformulation of the ODE to predict the next state of the current response based on previous state and the current voltage value.

\subsection{System Identification Method}
We implemented the SysID model by replicating the work in \cite{Subedi2024Aggregate} using 1 zero and 1 pole. This order structure was selected after manually experimenting with different combination of poles and zeros to select the simplest order that gives satisfactory accuracy at the minimum computational time. Similar to the B-spline-based model, transfer functions were generated for each of the 20 partitions. The SysID model, formulated as a non-linear optimization problem was solved using \textit{ipopt} solver. Details of the SysID modeling, algorithm and implementation can be found in \cite{Subedi2024Aggregate,SUBEDI2023100365}.

\subsection{Validation of Models}
The discretized Bspline-based model and the SysID models were both validated on the square step signal testing dataset. Each model was used to predict the current response, and their performances were evaluated against the actual measured values. This evaluation was performed using the  the Goodness of Fit (GoF) metric.

This is computed using the normalized root mean squared error (NRMSE), expressed as:
\begin{align}
    \text{GoF (\%)} &= (1 - \text{NRMSE}) \times 100, &
\end{align}
where NRMSE is given by:
\begin{align}
    \text{NRMSE} &= \frac{\sqrt{\frac{1}{n} \sum_{i=1}^n (y_i - \hat{y}_i)^2}}{y_{\text{max}} - y_{\text{min}}} &
\end{align}
The NRMSE adjusts the RMSE by normalizing it with respect to the observed range of the data, ensuring that its values are appropriately scaled.

\section{Results and Discussion} \label{sec:Results}
\subsection{Evaluation of Fitted Smooth Curve}
\begin{figure}[ht!]
    \centering
    \includegraphics[width=3.6in]{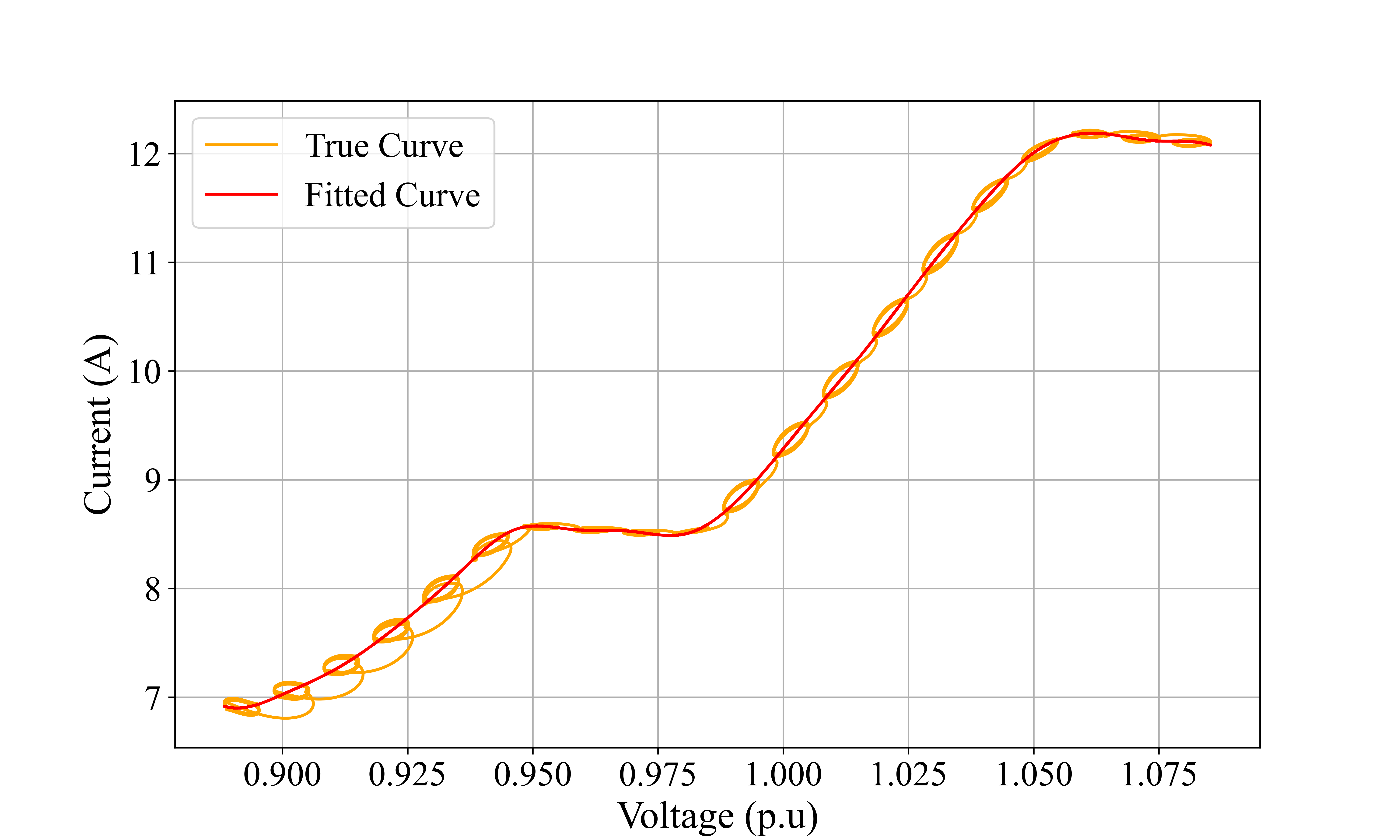}
    \caption{Approximated Smooth Curve Using B-spline Basis Functions}
    \label{fig:FDA_Curve}
\end{figure}

 In Fig.~\ref{fig:FDA_Curve}, the orange line shows the plot of the measured current response with Voltage variation. By using B-splines, this non-smooth curve is approximated with the smooth curve represented in red color. This smooth curve representation achieved a GoF of 98.70\%. The approximated smooth curve aligns with the piecewise representation in Fig~\ref{fig:Volt-Var}, showing the different regions of the current response to voltage variations. It however shows the non-linear behavior of the inverters response to voltage changes. This gives credence to the decision to adopt the partition model for the dynamics instead of the five region piecewise linear curve. Partitioning the curve into several smaller intervals guarantees better linearization than  the five region curve.  The B-spline approach successfully removed noise from the values of current response, making it easily differentiable over the voltage and also over time. With our focus on the system dynamics, the time derivative of the current response was computed and the coefficients of the first order ODEs in \ref{eq:regress_analysis} were determined for each partition. 

\subsection{Training Performance of FDA Model and SysID Model}
\begin{figure}[ht!]
    \centering
    \includegraphics[width=3.4in]{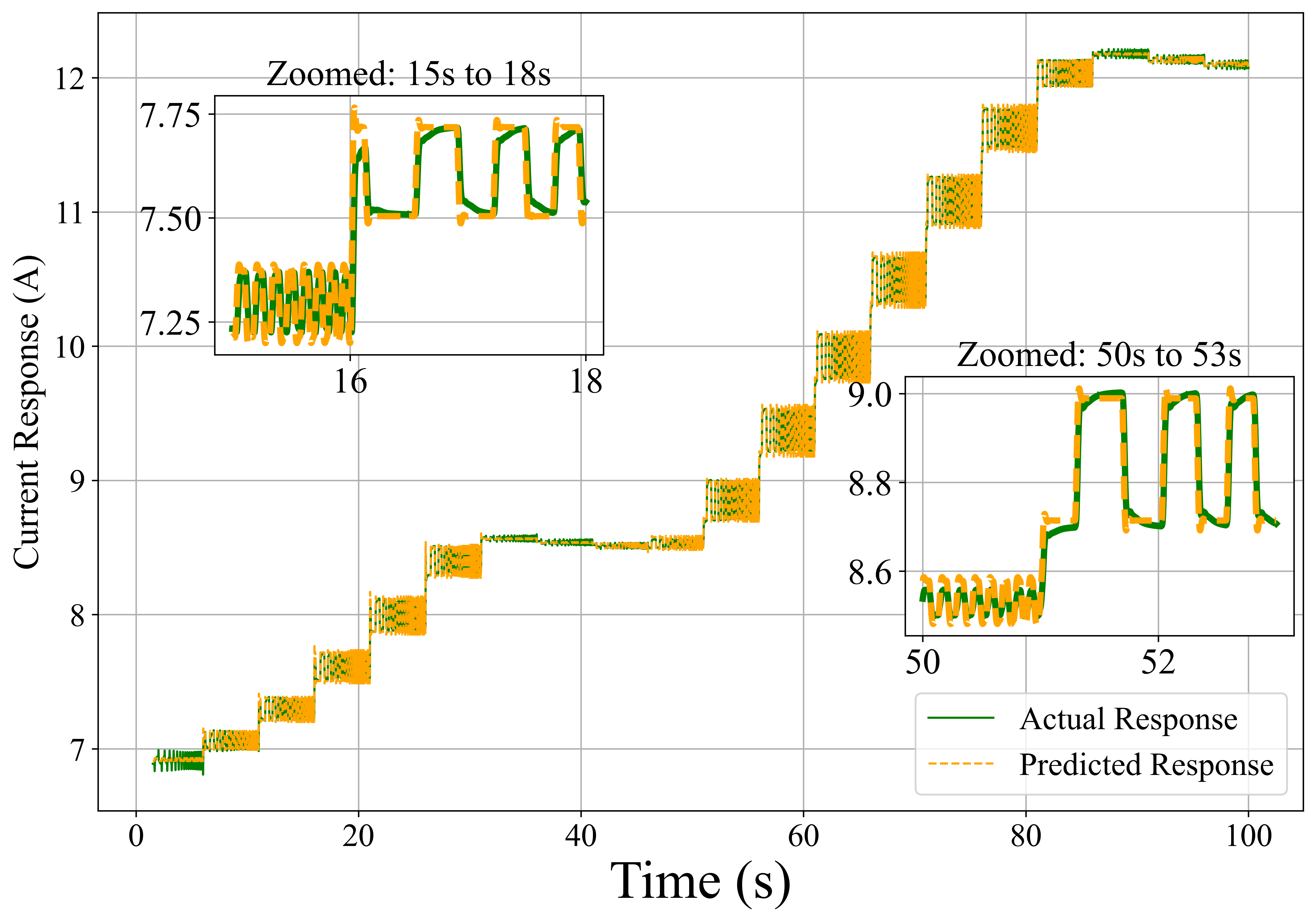}
    \caption{Training Performance of B-spline Based Dynamic Model}
    \label{fig:FDA_Training_Performance}
    \includegraphics[width=3.4in]{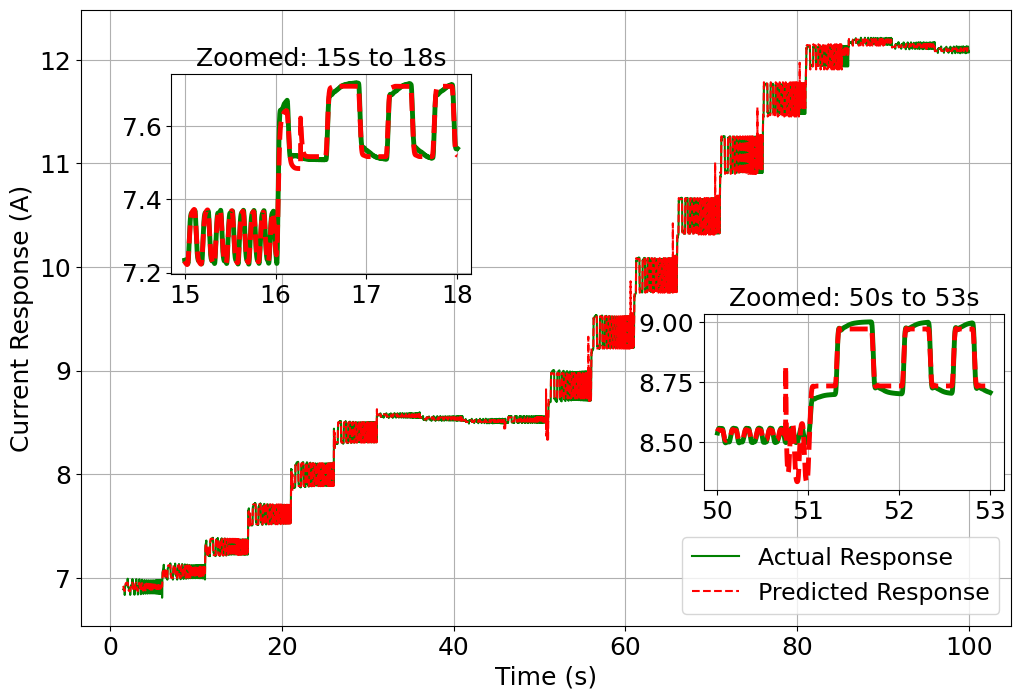}
    \caption{Training Performance of SysID Model}
    \label{fig:SysIDTrainig_Performancel}
\end{figure}

The extracted ODEs were used to predict the current response for each partition, and the results were aggregated to obtain the orange curve in Fig.~\ref{fig:FDA_Training_Performance}. The training result shows that our proposed approach succeeds at modeling the aggregate dynamics of the inverters response. The predictions by the extracted ODEs follow the trajectory of the true values of the current response with a GoF of 99.31\% and a computational time of 52.16 seconds. This compares well to the SysID model in Fig.~\ref{fig:SysIDTrainig_Performancel} which achieved a GoF of 99.65\% at a computational cost of 254.70 seconds. This shows that our approach significantly reduces the computational time and yet, achieves similar level of accuracy.

\subsection{Validation Performance of FDA Model and SysID Model}
\begin{figure}[ht!]
    \centering
    \includegraphics[width=3.4in]{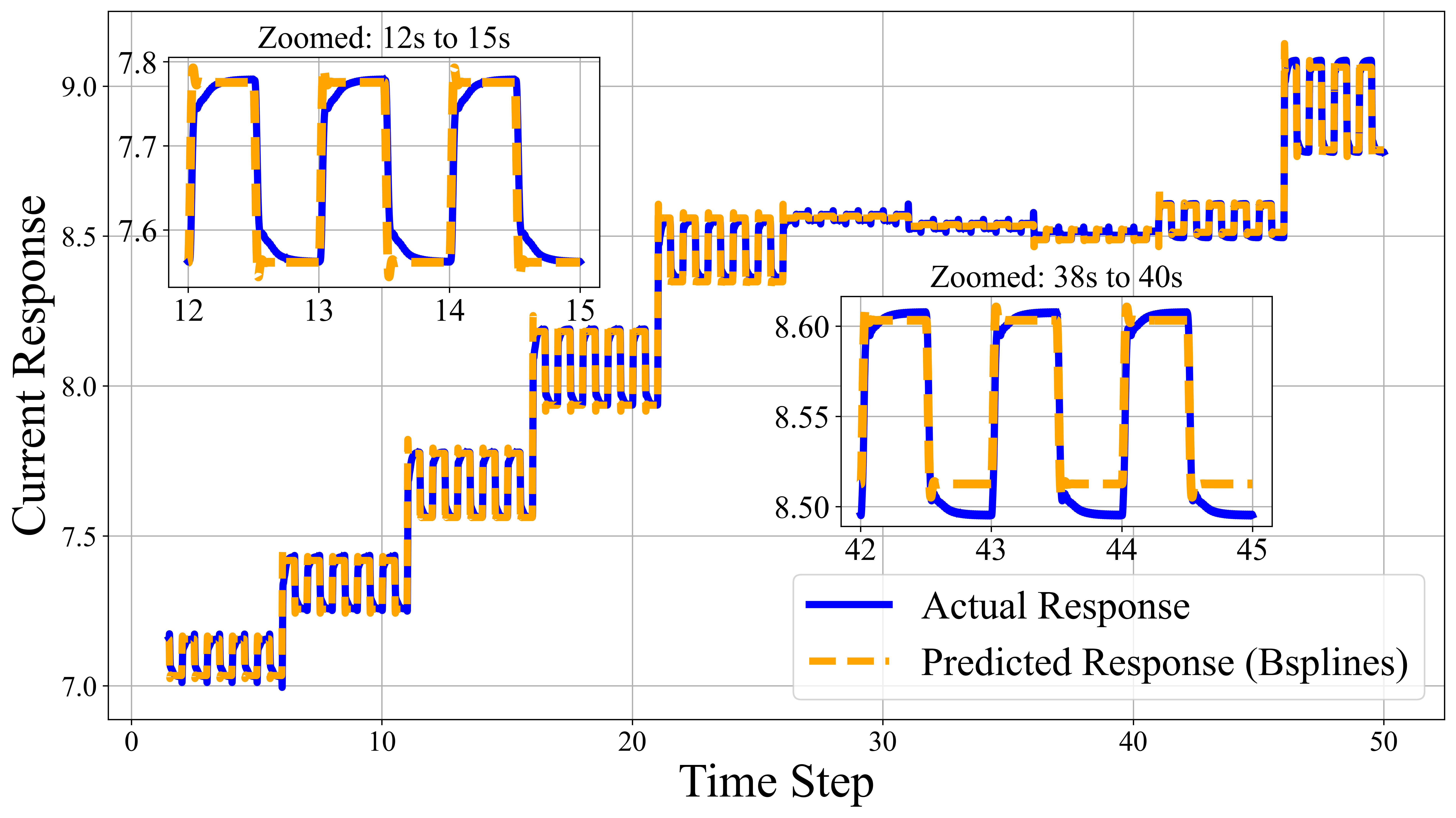}
    \caption{Performance of Bspline Based Model When Validated on Square Step Signal}
    \label{fig:FDA_Validation}
\end{figure}

\begin{figure}[ht!]
    \centering
    \includegraphics[width=3.4in]{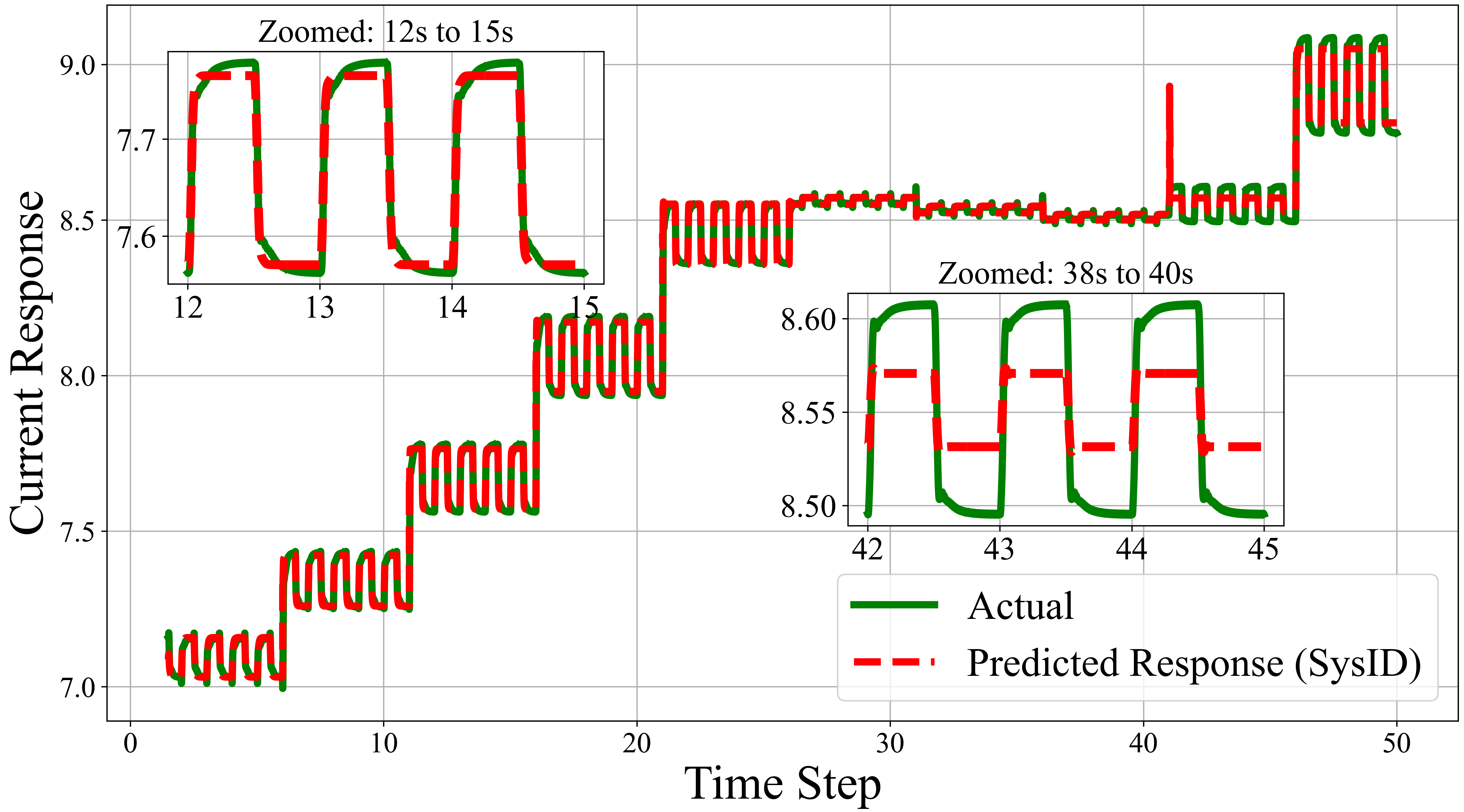}
    \caption{Performance of SysID Model When Validated on Square Step Signal}
    \label{fig:SysID_Validation}
\end{figure}

The validation of the two models on a different input signal, thus, square step signal shows the robustness of our proposed model to different disturbance patterns. The B-spline-based model performs well on the square step validation signal with a GoF of 98.74\%. Fig.~\ref{fig:FDA_Validation} shows how well the predicted current response follows the trajectory of the measured response even at the deadband. Similar to the training results, this compares well to the validated results of the SysID model which achieved GoF of 99.03\% as shown in Fig.~\ref{fig:SysID_Validation}. The validation results of both models were slightly lower than the training results. Although the SysID model shows a slightly higher GoF, the B-spline based model captured the deadband region better. The slightly higher performance of the SysID emerges from its non-linear programming formulations which enables it to capture the curves in graph a little better than our proposed model, but at a high computational cost. Even with linear programming formulation, our proposed model did a good job at capturing the dynamics of the system with a similar accuracy at a very low computational cost.


This observation was further reinforced when both models were repeated with higher orders. We tested both models for second, third and fourth orders. The number of poles for the transfer functions were changes from 1 to 2, 3 and 4. Similarly, we computed the second, third and fourth time derivatives of the current responses to model the higher orders. The results as shown in Table~\ref{tab:scalability_results} further proves the computational effectiveness of our proposed model over the SysID model. 

\begin{table}[ht!]
  \centering
  \caption{Higher order computational time of the two models}
  \label{tab:computational_time}
  \begin{tabular}{lrrrr}
    \toprule
    &
    \multicolumn{4}{c}{\textbf{Computational Time (seconds)}} \\
    \cmidrule(lr){2-5}
    \textbf{Model} & \textbf{1st Order} & \textbf{2nd Order} & \textbf{3rd Order} & \textbf{4th Order} \\
    \midrule
    B-spline Model &  52.16 &  52.95 &  53.25 &  54.07 \\
    SysID Model    & 254.07 & 318.53 & 952.07 & 1096.98 \\
    \bottomrule
    \label{tab:scalability_results}
  \end{tabular}
\end{table}

The scalability results also shows how the computational time of our proposed model only increases marginally with increasing order of the ODE. However, the SySID model increases exponentially with increasing order. This observation is due to the fact that our propose model eliminates the need to solve an ODE equations. Instead, the only additional task is the computation the higher order derivatives and the coefficients of the higher order linear ODE. In contrast with the SysID model, a higher order further complicates the non-linear formulation of the transfer functions, making it more difficult to solve.

\subsection{Discussion}
The B-spline based method offers a simplified approach to extract the system dynamics compared to the popular transfer function based SysID model. The B-splines' ability to approximate a smooth curve to represent nonlinear relationships provides a quick approach to extract the ODEs. The possibility to select the order and number of the B-splines polynomials also offers the advantage and flexibility to tune the model to accurately approximate the smooth curve. Furthermore, the model does not really require any training, but matrix multiplications to estimate the parameters of the splines, and that of the ODEs. The estimated curve is obtained by linear combination of the parameterized splines. This accounts for the improved computational speed observed. Conversely, the SysID model is formulated as a non-linear problem, with exponentially increasing computational time with increasing order.

\section{Conclusions} \label{sec:Conclusion}
In this study, we have presented an easier and simpler way of modeling the aggregate dynamic response of DERs dominated distribution system. Unlike the transfer functions based system identification model which utilizes non-linear programming formulation, our approach leveraged the simplicity of linear programming. By transforming the system data into smooth and differentiable functions, the time derivative of the DERs current response are easily computed. This enabled the use of least square estimation to determine the coefficients of a simple first order linear ODE for each partition of the dataset. Validation results showed that our approach produced similar accuracy as the SysID model but 4.8 times faster. The main limitation of this approach is the necessity to fit accurate smooth and differentiable function to the dataset. In cases where this is not possible, other dynamic modeling approaches such as the SysID model and neural ordinary differential equations are recommended. This study however, provides a quick and effective diagnostic tool to model dynamic systems for real time simulations and operations.


\bibliographystyle{IEEEtran}

\bibliography{Bibliography} 
\end{document}